\documentstyle[aps,aipbook,floats]{revtex}
\begin{document}

%\magstep2
\input psfig.sty

\rightline{OCIP/C-95-4}
\rightline{hep-ph/9505255}
\rightline{April  1995}

\title{The Measurement of Tri-Linear Gauge \\
Boson Couplings at $e^+e^-$ Colliders\thanks{Presented by S. Godfrey}}

\author{Gilles Couture$^*$, Mikul\'{a}\v{s} Gintner$^{\dagger}$ and Stephen
Godfrey$^{\dagger}$}
\address{$^*$D\'epartement de Physique, Universit\'e du Qu\'ebec \`a
Montr\'eal \\
C.P. 8888, Succursale A, Montr\'eal, Canada H3C 3P8\\
$^{\dagger}$Ottawa-Carleton Institute for Physics, Department of Physics,\\
 Carleton University, Ottawa, Canada K1S 5B6}
\maketitle

\begin{abstract}
We describe a detailed study of the process $e^+e^- \to \ell\nu_\ell
q \bar{q}$ and the measurement of tri-linear gauge boson couplings
(TGV's)
at LEP200 and at a 500~GeV and 1~TeV NLC.
We included all tree level Feynman diagrams contributing to the
four-fermion final states including gauge boson widths and
non-resonance contributions.  We employed a maximum likelihood
analysis of a five dimensional differential cross section of angular
distributions.  This approach appears to offer an optimal strategy
for measurement of TGV's.
LEP200 will improve existing measurements of  TGV's
but not enough to see loop contributions of new physics.
Measurements at the NLC will be roughly 2 orders of magnitude more
precise which would probe the effects of new physics at the loop
level.

\end{abstract}

\section*{Introduction}

A driving force behind high energy physics is the search for physics
beyond the Standard Model (SM).  An approach receiving
considerable attention generalizes  the
effects of new physics using effective Lagrangians (${\cal L}_{eff}$)
and tests for deviations from SM expectations.
While the fermion gauge boson couplings have been
measured to high precision by the LEP and SLC
experiments
%\cite{bigfit}
the vector
boson self-interactions have only just been experimentally verified
by direct measurement\cite{tevatron}.
Because the standard model makes precise predictions for the TGV's,
precision measurements constitute a stringent test of the
gauge structure of the standard model \cite{aihara}.
In this contribution we describe some
recent work on precision measurements of tri-linear gauge boson
couplings (TGV's) at $e^+e^-$ colliders; LEP200 and 500~GeV and 1~TeV
versions of the NLC.  A more detailed account of this work is given
in Ref. \cite{gintner}.

The processes $e^+e^- \to 4f$, including all tree level Feynman
diagrams that contribute to the same final state,
have been studied for a number of specific final states;
$\ell\nu \ell'\bar{\nu}$, $\nu \bar{\nu} \mu^+\mu^- $,
$q\bar{q} q' \bar{q}'$, and $\ell \nu q\bar{q}$.
Our present work consists of a detailed study of the last process,
$e^+e^- \to  \ell \nu q\bar{q}$, and its sensitivity to TGV's.  This
process can be fully reconstructed and
has the advantage of a higher branching ratio than the
purely leptonic processes while avoiding the ambiguities and
backgrounds associated with the purely hadronic modes.

There are three  effective Lagrangians commonly used to
describe TGV's which differ on the
degree of generality assumed \cite{aihara}.
\begin{enumerate}
\item {\it Most General Parametrization}.
The only assumptions here
are Lorentz and $U(1)_{em}$ invariance.
The parameters associated with the CP invariant
operators are
$g_1^Z$, $\kappa_Z$, $\kappa_\gamma$, $\lambda_Z$, and $\lambda_\gamma$.
$g_\gamma$ is always equal to 1 and in the SM at tree level
$g_1^Z = \kappa_V=1$ and $\lambda_V=0$.
Typically, radiative corrections from heavy particles will change
$\kappa_V$ by about 0.015 and $\lambda_V$ by about 0.0025.
%Although
%indirect limits from precision measurements exist \cite{burgess}
%they are not rigorous due to ambiguities so
The most robust limits
come from associated $W\gamma$ and $WZ$ production at the Tevatron
\cite{tevatron};
$-1.6 < \delta\kappa_\gamma < 1.8$, $|\lambda_\gamma | < 0.6$,
$-8.6 < \delta\kappa_Z < 9.0$ and $|\lambda_Z | < 1.7$.
\item {\it Non-Linearly Realized Goldstone Bosons} or {\it The
Chiral Lagrangian} includes custodial $SU(2)$ symmetry
which is experimentally verified to a high degree of accuracy.
The parameters of this Lagrangian are $L_{9L}$, $L_{9R}$, and
$L_{10}$.
$L_{10}$ contributes to the gauge boson self energies where it is tightly
constrained to $-1.1 \leq L_{10} \leq 1.5$ so we will not consider it
further.  $L_{9L,9R}$ are expected to be of order 1.
\item {\it Linearly Realized Goldstone Bosons} explicitly includes
the Goldstone bosons.
\end{enumerate}
Due to space limitations we only present results for the Chiral
Lagrangian but note that the different Lagrangians
can be mapped onto one another.

%The parameters from the two Lagrangians can be mapped onto each
%other:
%\begin{displaymath}
%\begin{array}{ll}
%g_1^\gamma   & = 1  \\
%\kappa_\gamma & = 1 + {1\over 32\pi^2} {e^2\over s^2} (L_{9L} +
%L_{9R} - 2L_{10})  \\
%g_1^Z & = 1 +{{e^2}\over {32\pi^2 s^2 c^2}} (L_{9L} + {{2s^2
%L_{10}}\over{(c^2-s^2)}})  \\
%\kappa_z & = 1 + {{e^2}\over {32\pi^2 s^2 c^2}} (L_{9L}c^2 - L_{9R}
%s^2 ) +{{4s^2 c^2}\over {(c^2-s^2)}} L_{10}  \\
%\lambda_\gamma & = \lambda_z
%\end{array}
%\end{displaymath}

\subsection*{Calculations and Results}

We studied the final states  $\mu^\pm \nu_\mu
q\bar{q}$ and $e^\pm\nu_e q \bar{q}$.  In the first case ten diagrams
contribute and in the second, 20 diagrams. We used the CALKUL
helicity technique to calculate the amplitudes and integrated the
resulting matrix elements using Monte Carlo methods to obtain the
cross sections and distributions.

For our numerical results we used the following set of
parameters: $\alpha=1/128$, $\sin^2\theta_w=0.23$, $M_Z=91.187$,
$\Gamma_Z=2.49$, $M_W=80.22$ and $\Gamma_W=2.08$.  We included the
kinematic cut $170^o> \theta >10^o $ where $\theta$ is the angle of
the charged lepton, quark, or antiquark relative to the beam axis.  We also
took $E_{\ell,q,\bar{q}}>10$~GeV.  In some of our results we
imposed that $|M_{(\ell\nu),(q\bar{q})}-M_W| \leq
10$~GeV where $M_{(\ell\nu),(q\bar{q})}$ is the invariant mass of the
$\ell\nu\; (q\bar{q})$ pair.

The object of the analysis is to maximize the sensitivity to anomalous
gauge boson couplings.  The longitudinal
components of the $W$ are most sensitive to anomalous couplings.
$W^-$'s ($W^+$'s) produced parallel to the incoming $e^-$
($e^+$) are dominated by transverse $W$'s while those in the
backward direction have a large $W_L$ content.  To extract the $W_L$'s
from the $W_T$ ``background'' we can use the $W$ decay products as a
polarimeter; $W_T$ decay products peak at forward or backward angles
while those of the $W_L$'s peak about $\cos\theta=0$ where $\theta$ is the
angle of the decay product with respect to the $W$ direction in the
$W$ rest frame.
The changing mix of $W_L$ and $W_T$
is illustrated in Fig. 1 by the quark angular distributions
for forward and backward $W$ scattering angles $(\Theta)$.
Additional information can be obtained by studying the decay
product azimuthal distributions which are subject to more complicated
interference.

\begin{figure}%[ht] % fig 1
%\vspace*{6.0cm}
% next line was used to print actual photo, commented out here.
% your syntax will probably differ.
\centerline{\psfig{file=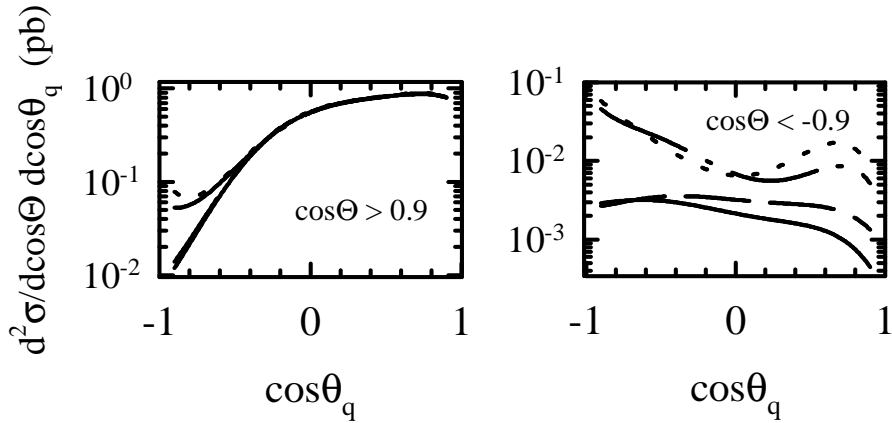,width=12.0cm}}
\caption[]{Angular distributions of the outgoing quark in the rest
frame of the $W^-$ in the process $e^+e^-\to \mu^+\nu_\mu q\bar{q}$.
The solid line is for the SM, the dashed line for $\kappa_Z=1.1$, the
dotted line for $\lambda_Z=0.1$, and the dot-dashed line for
$\lambda_\gamma=-0.1$.}
\end{figure}

To determine if anomalous couplings are measurable we use the
maximum likelihood method applied to the differential cross section:
\begin{displaymath}
{ {d^5 \sigma}\over{d\cos\Theta d\cos\theta_q d\phi_q
d\cos\theta_\ell d\phi_\ell } }
\end{displaymath}
where $\Theta$ is the scattering angle of the outgoing $W$'s and
$\theta_{q(\ell)}$ and $\phi_{q(\ell)}$ are the polar and
azimuthal decay angles of the outgoing $q$ ($\ell$) in the $W$ rest
frame.  We divided each of the angles into 4 bins.  Summing over all
1024 bins and comparing the SM predictions to those for anomalous
couplings we obtain the log likelihood function:
\begin{displaymath}
{{\ln\cal L} = \sum_i [-r_i +r_i \ln(r_i) + \mu_i-r_i\ln(\mu_i)]}
\end{displaymath}
where $r_i$ and $\mu_i$ are the measured and expected number of events
respectively.

We show the 95\% C.L. contours for the $L_{9L}-L_{9R}$ plane for
LEP200 and 500~GeV and 1~TeV NLC options in Fig. 2.

One can obtain additional information from the processes studied
here.  Possibilities we have studied but have not
included here for lack of space are:
\begin{enumerate}
\item {\it Initial state polarization.}  The distributions are different
for left and right handed initial state electrons mainly due to the
contributions of the neutrino exchange diagram to the $e^-_L$ mode
but not the $e^-_R$ mode.  This results in different dependences on
anomalous couplings adding to the measurement sensitivity.

\item {\it Single $W$ production.}  In the $e^+ \nu q\bar{q}$ (or
$e^-\bar{\nu} q\bar{q}$)  final
state, instead of imposing the kinematic cut that the final state
$e^+$ ($e^-$) be observed we can impose the cut that it not be
observed.  In this case the cross section is dominated by the
t-channel photon pole providing a mechanism for measuring the
$WW\gamma$ vertex independent of the $WWZ$ vertex.

\item {\it Off resonance production.}  The deviations from the SM value
cross sections can be dramatic off the $W$ resonances.  Although
the off-shell contributions by themselves
don't offer improvements to the on-shell $W$ results,
including the off resonance contributions can
improve the overall sensitivities in certain cases.

\end{enumerate}

\begin{figure}%[ht] % fig 2
%\vspace*{14.0cm}
% next line was used to print actual photo, commented out here.
% your syntax will probably differ.
\centerline{\psfig{file=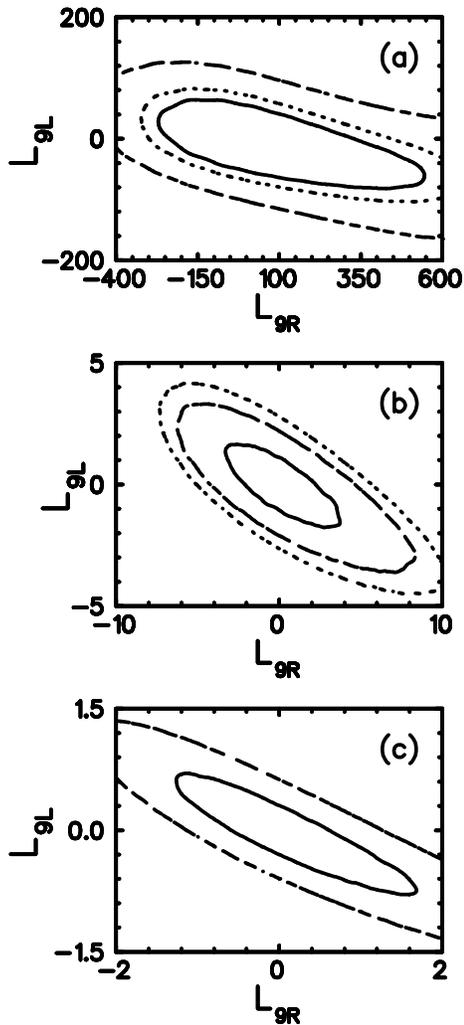,height=14.0cm}}
\caption[]{ 95\% confidence level sensitivity contours for $L_{9L}$
and $L_{9R}$ at LEP200 and the 500~GeV and 1~TeV options of the NLC.
In all cases the solid lines are the
sensitivities obtained by combining the $e^\pm$ and $\mu^\pm$ modes,
the dashed lines are for the $\mu^+$ mode only, and the dotted line
is for reduced luminosity and combining the 4 modes.
(a) $\sqrt{s}=175$~GeV: There are no cuts on $M_{\ell\nu}$ or
$M_{q\bar{q}}$. The solid and dashed lines use L=500~pb$^{-1}$ and
the dotted line uses L=300~pb$^{-1}$.
(b) 500~GeV: Include the cuts $|M_{\ell\nu (q\bar{q})}-M_W|<10$~GeV.
The solid and dashed lines use L=50~fb$^{-1}$ and
the dotted line uses L=10~fb$^{-1}$.
(c) 1~TeV: Include the cuts $|M_{\ell\nu (q\bar{q})}-M_W|<10$~GeV.
The solid and dashed lines use L=200~fb$^{-1}$. Here the reduced
luminosity contour of L=50~fb$^{-1}$ lies on top of the $\mu^+$ curve.}
\end{figure}

\section*{Summary}

We have presented some results from a study of TGV's
in the process $e^+e^-\to \ell \nu q\bar{q}$. We
employed a maximum likelihood analysis of a five dimensional
differential cross section of angular distributions.
LEP200 will improve on existing measurements, but not sufficiently
to observe deviations originating from loop contributions from heavy
particles.  On the other hand the NLC will be able to measure these
couplings to better that a half of a percent which would be
sensitive to radiative corrections to the TGV's.

\acknowledgments

SG thanks the organizers of TGV95 for the invitation to attend a most
enjoyable meeting and the Deans of Research and Science at Carleton
University  for financial support to attend the meeting.
This research was supported in part by NSERC Canada
and Les Fonds FCAR du Quebec.

\end{document}